\documentclass[conference]{IEEEtran}
\usepackage{url,setspace,amssymb,tikz,algorithmicx,algorithm}
\usepackage[numbers,sort&compress]{natbib}

\usepackage[OT1]{fontenc}

\title{proposal}
\begin{document}
\title{UniqueID: Decentralized Proof-of-Unique-Human}
\author{\IEEEauthorblockN{Mohammad-Javad Hajialikhani}
    \IEEEauthorblockA{Department of Computer Engineering\\
        Sharif University of Technology\\
        mjavadalikhany@gmail.com}
    \and
    \IEEEauthorblockN{Mohammad-Mahdi Jahanara}
    \IEEEauthorblockA{Department of Computer Engineering\\
        Sharif University of Technology\\
        mmjahanara@gmail.com}}
    \maketitle
    
\begin{abstract}
    Bitcoin and Ethereum are novel mechanisms for decentralizing the concept of money and computation. Extending decentralization to the human identity concept, we can think of using blockchain for creating a list of verified human identities with a one-person-one-ID property. UniqueID is a Decentralized Autonomous Organization(DAO) for maintaining human identities such that every physical human entity can have no more that one account. One part of this identity is simply the user's claim on one of his unique, permanent, and measurable characteristics -biometrics. Blockchain has proved its integrity as a platform for storing and performing computations on such claims. The biggest challenge here is to ensure that the user has submitted his own valid biometric data. Human verifiers can check if there is any inconsistency in other users' data, by peer-to-peer checks. For preventing bad behavior and centralization in the verification process, UniqueID benefits from novel governance mechanisms to choose verifiers and punish unjust ones. Also, there are incentives for honest verifiers and users by newly generated tokens. We show how the users' privacy can be preserved by using state-of-the-art cryptographic techniques, and so they can use their identity without any concerns for votings, financial and banking purposes, social media accounts, reputation systems etc.
    
\end{abstract}
\section{Introduction}
Identity verification has always been an important problem in online systems. While identity is a key to solve a wide range of challenges, including Sybil attacks \cite{Sybil} and accountable votings, there is no universal system for holding human identities. Identity databases are usually created and managed by governments or financial institutions privately. Social media and messengers solved the problem partially by relying on other systems such as E-mail, mobile phone number, credit card information, and other reputable social media accounts. These methods are not suitable for highly critical applications(e.g. voting) and prone to violating users privacy. Currently, there is not any system for assigning every online human exactly one account, which can be used for logging into other systems as well.

Centralized models for identity verification are all prone to misuses, fake identities, and political conflicts. Identity is a political concept, and nation-states will come against every system that aims to collect their citizen identities. Moreover, we need to make sure that such an authority does not create fake identities, or excluding a specific group of people from the system in favor of itself.

Blockchain brings us a decentralized way for recording specific data in an immutable ledger -which the \textit{correctness} of data is also guaranteed by computer code, based on pre-specified rules. Decentralized money is a direct application of such a ledger. A monetary system can be reduced to statements in forms of \textit{``X has c coins"} and \textit{``X gives Y c coins"}. Bitcoin \cite{bitcoin} just records these statements -\textit{transactions}- in its Blockchain. Nature of identity verification is very close to this \textit{record} logic because identity is simply some information about you, and some claims about the correctness of those information by other people.

UniqueID aims to bring online and decentralized one-person-one-account system based on biometric data for recognizing people, human verifiers for identity verification processes, and blockchain for recording data. Similar to Bitcoin which incentivizes people who contribute their computational power to network with new Bitcoins, UniqueID uses a token system and a novel governance mechanism to incentivize human verifiers and distributing operations while preventing the creation of a monopoly of power and bad behavior.

Recently there have been efforts in order to create decentralized identity verification systems with different approaches. \cite{proofofpersonhood, uport, sovrin} are examples of such systems. To the best of our knowledge, among all the systems which provide a unique account for every person, none of them are permissionless, fully decentralized, and permanent at the same time.

Identity is also an important problem, not just in the digital world. Roots of many nation-wide issues are identity-related issues. Over 1 billion people do not have any identity document, and can not participate in financial and legal transactions \cite{worldbankid}. Old institutions for democracy are changing too slowly and are not efficient, because of bureaucratic and paper-based identities leads to costly votings \cite{democracyearth}. Universal basic income, as a novel idea, can be executed only in the existence of a worldwide identity system \cite{basicincome}.


\section{ Problem Specification}
In this section, we define our problem formally.

\subsection{Problem Statement}
The simplest formulation of our problems is: \\
\textit{Create a decentralized system for assigning exactly one public key to every human, without violating users' privacy. }

This definition may create some debates. First, we must define words like user, human, and decentralized as we see in next section. We can see this system as maintaining and managing a list of valid public keys without trusting any centralized authority. By \textit{system}, we mean a socio-economic entity where some group of people, encouraged by economic incentives, form a specific structure toward achieving a goal. 

\subsection{Terminology}

\subsubsection{Human}
When we want to design an identity system for humans, we must tackle two questions:
\\
1. What makes an entity a \textit{human}?
\\
2. What makes this human persistent over time? How can we say one person is the \textit{same} over time?
\\
The first question is crucial, to determine \textit{who} can enter the system. Can we name a strong Artificial Intelligence a human? It seems that all of us have some consensus on a definition of a human. It's a combination of some specific body properties (e.g. shape, face, etc.) and intellectual abilities (e.g. memory, language, etc.). 
\\
The second question is harder, and a famous unsolved problem in philosophy \cite{identity-personal}. If we want to map the identity of a person to any of his properties, we lose something. for example, if we define personal identity over time as a person's body, it is known that all body cells are replaced after some time. But we just need some properties that are persistent over time for most people (without any fundamental philosophical reason). We can name three of these properties: 
\begin{itemize}
\item{Biometrics:}
Some specific human body structures have proven to be persistent and unique over time. Fingerprints, DNA and iris scan are examples of proper biometrics.
\item{Memory:}
The average human brain can memorize some data and recognize it over time. Persistence of memory is almost true for all people.
\item{Trust:}
This is less clear than the previous properties. Trust is how other people know and recognize one person. Basically, it is a combination of some mind-body properties. However here, \textit{judging} over these properties is up to others. Trust is how other people see someone as persistent and unique over time. Human judgment can make some errors, but it works well in many situations.
\end{itemize}

\subsubsection{Decentralized}
Defining decentralization, especially in a rigorous form is very hard mostly because it takes different forms in different contexts \cite{meaning-of-decentralization}. We focus on decentralization in its political form here: there is no specific entity -state, corporation- or person who can affect the system, creating or removing identities. One side of this decentralization is the inability of an entity in changing ledger information: account balances, biometrics etc. Blockchain already makes this part possible. Another side is the inability in adding identities and influencing how people trust each other, and here is where our design should gain such decentralization.
\\
We always see the first side as \textit{``How much money or coordination is needed for attacking the system"}, e.g. Bitcoin hashrate gives us a minimum on computational power (that can be expressed in computational devices that one can buy with money) that an attacker needs for the double-spending attack. Besides, how current hashrate distributed shows us another aspect of decentralization: minimum number of entities that form at least 51 percent of the system. However, we have harder times dealing with the second side. One can formalize that as \textit{``How many people should collude, and how much money they need, in order to do X in the system?"} X can be making one fake identity, making infinite fake identities, preventing someone from entering the system (censorship) and any other unwanted incident. We analyze our system from this point of view in section 6.

\subsubsection{Incentives}
We can see this kind of system as a form of the cooperative game, where people interact with each other for reaching a shared goal. For this reason, there must be enough incentives for parties to participate. Formally speaking, we should determine the amount of effort, energy and other resources every party has to put, and how many tokens he gets back. Analysing this market can give us the equilibrium and the prediction of what will happen at the end. In the case of Bitcoin, e.g., there are always incentives for some group of people (miners) for running full nodes and miner equipment for earning new Bitcoins, so the system will be alive as long as these coins are paid to them.

\subsubsection{Privacy}
Privacy is the ability of an individual or group to have control over all the information about themselves and to whom they express. 
Privacy and data protection are of the most important challenges in every blockchain-based system because it is known that decentralized systems are inherently transparent, and all data can be accessed by any node. As shown in recent researches, many cryptocurrencies are at the risk of linking transactions and other data leaks. This transparency and lack of privacy is a cost of high security and immutability of them. We need more consideration for preserving privacy in a permanent identity system which can be used for critical purposes. Cryptographic advancements, such as functional cryptography, program obfuscation, and ring signatures can be used for enhancing privacy in UniqueID. We will discuss privacy concerns and imaginable solutions in depth in next chapter.

\section{UniqueID Design}
In this section, we come up with a detailed explanation of how UniqueID works and solves the decentralized proof-of-unique-human problem. First, we provide an overall flow of the system, and then we cover the concepts and components which our system consists of.

\subsection{Overal Flow}

Here we focus on the flow of processing one user's identity for understanding how UniqueID works.  Users should submit their biometric data to the blockchain and then prove the validity of the claimed biometric. This proofs are “social” and based on trust; It means you must get some “biometric reproducing certificate” from some of the previously accepted users, i.e. verifiers. These verifiers ask you to “reproduce” your claimed biometric in presence of themselves. After collecting enough trust, the user is recognized as a valid and unique one. At a technical level, Blockchain brings us the storage and computation we need for these tasks.

There are some questions arising here, such as how users get assigned to verifiers, what are the incentives of verifiers for checking others' claims and where is the starting point of the system. We discuss these questions and provide solutions for them in following subsections.

\subsection{Biometrics}
Biometric data is how UniqueID identifies and recognize ``humans". We need almost permanent, unique and well-studied biometrics for our system that has an acceptable accuracy level, and also storing them on blockchain do not create any privacy concerns.
Biometric data will be used to check the uniqueness of individuals and prevent the creation of fake identities and Sybil attacks. A wide range of physical and behavioral biometric indicators have been developed during last decades in the literature of biometrics-based identification, including iris, face, finger vein, hand geometry, typing, etc \cite{intro-biometic}. Advancements in machine learning and image processing offer many well-studied and tested methods of biometrics-based identification with equal error rate (EER) of at most $0.01$, let alone methods with much higher accuracy. Roughly speaking, compounding at least 4 of uncorrelated different biometrics as a biometric indicator, our system is ready to serve and identify 100 million people.

\subsection{Trustless Setup}
In a system where current users should approve new ones, one should come up with a mechanism for approving very first users. One method is beginning with a sufficient number of trusted persons and introduce their identities (biometrics and keys) as verified users. The challenge is how to select these persons without any need for a central third party? Below we explain three solutions for this issue.

\subsubsection*{Transparent Peer-to-Peer Setup} 
Transparency can be the key to solving the trustless setup problem. Usually, there is a community of enthusiastic people around these kinds of projects. By providing relevant information in a public channel, the community invited for joining a public party in special place and time. Everyone who aims to join has to send her proposal including name, biometrics and some identity document -just for transparency purposes. In the party, they start to verify each other in a peer-to-peer manner. Because the party is public, everyone can join and there would no monopoly or closed group of individuals at the beginning. Also, it's near impossible to fake an identity because it can be detected and claimed by just one person in the party. There can be a live cast of all of the steps of the party and all of the relevant information in a public channel for forwarding transparency.

\subsubsection*{Trusted Famous People}
There are some people with a public profile who have no incentive for sabotaging the system, e.g. university professors, known artists, and athletes or popular cafes. This is usually because of their high reputation among other individuals. Collecting some of these people who want to join in early verifying, and broadcasting all the information and profiles can be a starting point for the system.

\subsubsection*{Decentralized CAPTCHA Party}
CAPTCHAs \cite{CAPTCHA} are mechanisms for recognizing humans from machines. CAPTCHAs are AI-hard and it takes reasonable time for humans to solve them. Consider we construct some hard CAPTCHA that takes about 2 minutes for an average person to solve. We then organize a CAPTCHA party: in a specific time around the world, based on some common randomness (e.g. blockhash) a predefined smart contract starts assigning CAPTCHAs to participants. People can prove their uniqueness by submitting an answer to the assigned CAPTCHA before the party ends.

\subsection{Verification Mechanism}
We assume that there is a number of verifiers, say 100, in the city. Location and city of every verifier recorded in the blockchain. Call the user's public key $pk$.

 after one user enters the system with his biometric, he should verify his identity and get certificates from a sufficient number of these verifiers(3 can be enough here). Assigning verifiers to users would be done via blockchain regarding user's location and some common randomness. The randomness can be produced by another smart contract, with methods like  \cite{rhound}. 
  This randomness,$ R_i$ combined and hashed with user's public key ($SHA256(R_i || pk$) can determine the assignment.

The user then goes to all of the assigned verifiers in person and \textit{reproduce}  his biometric in the presence of verifier. Reproducing means sampling user's biometric with verifier's biometric device, and then ensuring equality of this sample with previous biometrics on blockchain with some local computation on verifier's device. If the verifier ensures that two biometrics are the same, he $accepts$ the user and issues a certificate with his signature on the blockchain. This process is for preventing fake biometrics attacks and identity thefts from happening because verifiers can make sure that the biometric is real and belongs to the person who claims it.

After user gaining these certificates, the system automatically accepts him as a valid and unique person. If only one of the verifiers does not issue the certificate, the user can ask for a reassignment. This is possible when one of the verifiers is corrupted or has made some error. There is no incentive for a verifier in rejecting a real user, as the user can ask for a reassignment afterward.

For preventing Sybil attacks, and ensuring that an adversary with many attempts can not break verification process and assigned to his favorable verifiers we add a barrier in entering the system. There are just three ways you can ask for the initial assignment:
\begin{itemize}
\item Via $invitation$. Every verified user has two invitations that can give them to other public keys so they can submit their biometrics and start the verification process. because there is a limited number of verified users, there will be always a limited number of verifications (and therefore performing an effective attack would need large human resources).

\item Via $Stake$. The user can put some specified amount of money at stake for starting the verification process. At the end if he verified by the system, the system automatically will return his stake, otherwise, his stake will be locked forever. The amount of stake should be in a direct relation with all unverified stakes in the corresponding city, means the more unverified users, the more stake a user should put.

\item Via $Verifiers$. Every verifier has an authority of starting the verification process for a specified number -according to his reputation and trust- of new users.
\end{itemize}

Also, the process of assigning the verifiers should be done respectively, means the user just know who is the \textit{next} verifier. If the user completed and verified with the current verifier, then the next one will be assigned by the system. This should restrict the power of an adversary in rejection of an unwanted assignment.

\subsection{Trust Delegation}
Trust is the fundamental key to bring healthy behavior in our system. Every unique person has one ``trust" token which can delegate it to everyone. With this non-tradable token, one can delegate his verifying authority to someone else, because he can not actively verify new users. Thus the verifier which possess a greater number of trust tokens have more verifying power, and more the system can trust their acceptance/rejection.

The system will then give verification power to those which possess at least $x$ trust tokens, in order to make verification process more reliable. this parameter must set based on the city population and maturity of the system. Verifiers can collect trust from users by Transparency. For example, a verifier can provide access to a live cast of the verification processes on a youtube channel, or permits everyone coming to his office (doors are open!). By providing more transparency, more people can trust him.\\
Also, every user must have incentives for delegating her trust to a trustworthy person. Restricting some important features of the system (e.g. Universal Basic Income) from those who did not delegate their trust is one solution. Also if a corrupted verifier identified (there will be some mechanisms for identifying corrupted behaviors), all his trust will be suspended for some period of time.

\subsection{Native Token}
We introduced a native token for our platform, However, For every  token system we should answer to at least 3 questions:
\\
1) What is the necessity of creating a new token, Or why the system needs it?
\\
2) What is the utility of this token for users, Or why it has non-zero value?
\\
3) What is the monetary system, Or how tokens issued and distributed?
\\
\subsubsection{Nessecity}
First, we need a token for creating incentive mechanisms. every decentralized platform needs sophisticated and well-studied incentives, for encouraging healthy behavior and punishing bad actors. One of the powerful tools to design mechanism is money because it is related to person's desirability and profits. In our system verifiers rewarded by tokens every time they verify a new, fresh user. Also, new users get some tokens for the first time they enter the system. Also, some important applications, such as UBI, need some form of currency.

\subsubsection{Utility} 
The most obvious use of such a token is for payments and store of value. Incentives bring new users to the system, and new users bring more acceptability and a wider market, projected in token price, bringing new incentives for further users. This positive network-effect, combined with a mechanism for more incentivizing first movers(but not make them giant Whales!) can bring non-zero value for the token. Also, this system can host many other applications which need a unique and permanent identity, such as digital banking, property and real estate on the blockchain, reputation systems etc. These systems can provide their services on the UniquID blockchain, using a native token or at least paying fees for their smart contracts.

\subsubsection{Monetary system}
Our goal is to achieve an issuance scheme which is fair and keeps the value of the token. Our proposal is selling fix amount of tokens (say, $a*x$) in pre-sale (with an Initial Coin Offering or similar methods) and then issuing $x$ tokens for every new user. This tokens distributed between verifiers of that user and herself. So after $a$ new users, tokens in the hands of users is equal to ICO tokens. This can provide $fairness$ for the system because initial holders are not giants anymore, and after the system became popular(=more users) their influence gets reduced. Also, initial investors have incentives for investment because they can buy cheap tokens in early stages. It will be promised to them that their tokens have an influence equal to $a$ users, and they can guess what they earn by the estimated value of an identity system with $a$ users. This issuance scheme is linear in the number of users, However, the utility of users(network effect) is quadratic, so we can predict a slight rise in price over time. At least, It's not a super-inflationary system: More user's, more utility, more demand and more supply responding to them. Choosing $a$ parameter and schemes for incentivizing early users (e.g. changing reward over time) is for further research.

The figure for comparing network effect and money supply.

\subsection{Decentralized Governance}
There is no independent entity for decision-making in decentralized systems, so they face governance problems on many levels. Long discussions on  DAO fork and Bitcoin Segwit2x proposal are clear examples of this issue. In UniqueID, There is a new concept that should be governed as well, human work. There should be some mechanism for detecting and punishing \textit{corrupted} verifiers.

UniqueID is inherently one-person-one-vote, and this gives us new opportunities for shaping our governance model. We introduce some hierarchical representative democratic system, based on three layers of representatives.

First, small communities of $50-100$ persons choose one person from themselves. They put their trust and vote on him, and can also track his activity and ask him about that. Because they are a small group of people, it is expected that they know each other well and can make necessary communications between themselves and choose a qualified representative. these are layer-1 representatives, and they should know each other and follow system changes and discussions, and also choose layer-2 persons. Every $30-40$ of these layer-1 representatives should gather and choose one representative. And the same for layer-3: every $20-30$ layer-2 people can choose one layer-3 representative from them. 


This system should decide for critical decisions such as choosing parameters and consensus rules. Majority consensus seems a very good method for making decisions in a system with verified identities but also can cause some problems. Not \textit{every} decision can be made by a majority (51 percent) of votes. Changing some critical parameter of a system (number of verifiers needed for verifying one person) is a clear example of this. So we need to group decisions based on their importance and put some thresholds on the percentage of votes needed for making decisions in every group (super-majority rules).

Because of low participation, lack of incentive for voting and lack of knowledge for making correct decisions we should give some authority to representatives. But for lowering corruption and collusion, in addition to transparency, every decision needs greater consensus among higher layers. for example, if we need 51 percent of all user votes for some parameter change, we need 68, 85, 95 percent of votes for layer 1,2,3, respectively.

In this process, people can track their trustee activities and votes, and ask for explanation and reasons. Everyone can change his trustee every time, and delegate his trust to another one. However, representatives should not lose their position with leaving one of their trusts, so there can be a threshold(e.g. 20 percent) that when they leave, the person becomes invalid.


\subsection{Implementation}
Formally, UniqueID is a decentralized autonomous organization (DAO) which its operation is based on a decentralized IT infrastructure. In this section, we focus on the IT infrastructure part. While there are various possible scenarios of implementation for UniqueID, ranging from building on top of existing blockchain-based platforms to the creation of a standalone blockchain, we describe a modest implementation scenario using existing technologies.\\

UniqueID uses a set of smart contracts on Ethereum blockchain as a \textit{compuational back-end} and \textit{immutable storage}, all the functionalities of system and interactions of users (e.g. claiming a new identity, verification process, delegation of trust, etc.) are implemented in solidity and get done on-chain. Moreover while UniqueID stores critical information like the state of users in the verification process and the hash of biometrics data (as a unique identifier) on the blockchain, in order to reduce the prohibitively large cost of on-chain storage, biometrics data will be stored on Swarm. Each identity holder is responsible to make sure her biometric data is always available on Swarm, it means that in case of data loss she has to upload same biometric data to the network again otherwise her identity get temporary invalid.

One of the core functionalities of UniqueID is to compare different biometric data in order to verify the uniqueness of identities and verify consistency of reproduced biometric data by new identity claimers. While it's possible to give this responsibility to human verifiers, in order to minimize human mistakes while maximizing automation and transparency, UniqueID uses Truebit \cite{Truebit} protocol to get this task done. 

\subsection{Improvements over verification}
The verification system we purpose face some challenges. Bribing or coordinating some verifiers for creating fake identities is the most important one. Fake-ID prevention is the only thing we design our system for solving that. \\
We can see two directions for solving this attacks. First, Set some sophisticated and hard rules for entering people that guarantee hardness of attacks. Second, design some mechanisms for detecting and removing fake identities and punishing bad verifiers. In this section, we explain a mechanism of the second kind. \\
We first design a very secure registration way that there is zero-probability in entering with a fake identity, calling it A-judge. This can be a very hard process, including being verified by all verifiers in a specific city. There can be gathering of verifiers every 6 months in a public place, and handling all A-judges there.

Using A-judges, we can design a system for lowering amount of coordination between verifiers for creating fake identities. We give every verifier and every layer-3 representative limited number of \textit{Identity re-checks} per month. They can use these re-checks for calling specific identities (here, biometrics) forgoing through A-judge. These called identities should pass A-judge in a specified time, Otherwise, they lose their identity. Also, they have great incentive for going through this judge, because they rewarded by tokens. This way we can find fake-identities and punish bad verifiers.

Every verifier stakes X tokens and locks them. If the system determines some verifiers as a bad verifier, his stake goes for anyone who called A-judge for his fake identity. So verifiers have the incentive for calling A-judges over fake identities if they know any. This makes coordination very hard: even if you bribe heavily some verifiers, After that, they can call A-judge over that identity. There should be also some random checks for preventing other forms of bad behavior.

\subsection{Privacy}
We can see different aspects of data privacy concerns in UniqueID. Below we discuss each of them and their solutions in more detail.

\subsubsection*{Biometric Data}
In UniqueID we need user's biometric data for the human identification process. Leaking biometrics can lead to the harmful use of them, such as proving procedure in courts or searching for a possessor of a special account by authorities. What we need is a way for preserving the functionality of the system, without exposing biometrics to anyone.

More formally, the process of biometrics can be described in terms of user $U_i$ puts his biometrics $B_i$ in a ledger $L$. This ledger then compute the function $AI(U_i,U_j)$ for every $j \leq i$, ensuring $U_i$ is a fresh and new biometric. The output of the $AI(x,y)$ can be binary, or a number in [0,1] demonstrating the similarity between x and y. We want to preserve the functionality of $AI$ function, But keeping $U_i$ secret. 

Finding an efficient and decentralized solution for this is an interesting challenge. We present a solution here which needs trusted setup. Consider there are public and private keys $pk,sk$, where $pk$ is known to the public but nobody knows $sk$. Also define $PrivAI(X, Y)$ as a code that first decrypting $X$ and $Y$ with $sk$, and then computes $AI$ on them. $ObfPrivAI$ is obfuscated version of program $PrivAI$ (Indistinguishable Obfuscation). 

In a trusted setup, we can compute $sk,pk,PrivAI$ and $ObfPrivAI$ in one session, and then removing $sk$ and $PrivAI$. Then we put $ObfPrivAI$ on the ledger instead of $AI$. Every user sends an encrypted version of his biometric, $Enc_pk(U_i)$, to the ledger. Now we can compare biometrics and gain predefined functionalities, without exposing user's biometrics. Designing some Multiparty setting for trusted setup, Proving the correctness of trusted setup procedure and thinking about more efficient protocols can be further challenges. 

\subsubsection*{Secure Logins}
One can use UniqueID blockchain for designing some decentralized application, especially as the one-person-one-account property is attracting for some purposes, such as social media or banking. It is plausible that these systems should hold user's data private, and the identity of users should not be revealed. More formally, there should be no relation between user's public key (or biometrics) and his social media account. This can be solved by introducing anonymous tokens, means having a special token (one token per person, specified in application's smart contract) for logging in. Adding the ability to send tokens anonymously between accounts, the user can send his app-token to anonymous account and start with a fresh identity. This ability can be achieved by ring signatures \cite{RingSig} or zero-knowledge proofs \cite{ZeroKnowledge}, working examples are  Monero, Zcash or Raiden network. There should be also a way for everyone for burning his previous identity (by revealing the relationship between his main and application account) and start with a fresh identity in the application. This process is crucial for preventing the creation of the token market.

\section{Challanges}

\subsection{Geographical Dispersion}
Until now we assume that there is one city, where all verifiers and users are close to each other and interaction between them is easy. In the real world, however, we have thousands of cities in different countries, villages, no-residence areas etc. This variety makes some serious challenges for a system relying heavily on peer-to-peer verifications in the same place.

The first problem is how new cities and places will join the system? So far we design our system and ``first verifiers" based on assumption that they are all in the same place, That's not true in real world. Imagine a scenario where we start the system from Tokyo, and after a while, people in Vienna want to join the system. Should they go to Tokyo and Verify their identity?

Some possible solutions we talk about them here, but solving this problem needs more thinking. First, we can have another trustless setup in the new city. But for keeping the integrity of the system, other existing verifiers around the world should take place in this setup. The system should choose these people and provide their expenses in a secure and decentralized way (Making a data feed for the list of valid cities, and some kind of voting or random assigning between volunteer verifiers). So based on these new verifiers, users in this city can start their verification processes. Another solution is to make as many as verifiers in the first setup, Bringing people from different locations around the world in one borderless place (an airport, e.g.) and then start the system with these verifiers.

Second is how residences in small cities and villages can join the system? Areas with population less than 10000  may don't have any verifiers at all. One possible solution is assigning them to nearest cities, But it makes an entry barrier for them. Another solution is sending some verifiers from near cities for verification, But it can bring some attack vectors and unpredictable behavior. 

\subsubsection*{Other methods}
It seems that using other methods of verification, we can solve the identity problem without facing geographical dispersion. We think that this challenge is fundamental, i.e. near every system for securely verifying identities has this problem.

First, verifying human identity always require some human work. Until now, It's impossible for an AI for answering the question of ``sameness" of two human objects over time. Nature of identity, Combination of physical and mental attributes, is the root of this hardness. So people should identify people. Now, we have choices of making this verification in same physical place, or from long distance. We argue that long-distance methods are prone to errors and attacks: Every long distance method consists of sending some documents and information and making some interaction (video chat, voice chat etc.) between verifier and user. Both have limitations: every document (video proof of your biometric, voice captcha etc.) can get hacked with AI methods. Even further, AI enables us to fake a face in real-time video chat.

Geographical dispersion of people around the world makes some fundamental challenge for identity systems. Even in a simple(and bad!) system which you should get some certificates from your friends (and you get verified after that) we face this problem because people you know are mostly near you.

\subsection{Hard Forks}
\subsection{Massive Computation and Storage}
The current design of UniqueID requires permanent storage of all claimed biometrics and also a computational infrastructure for automatic uniqueness check of new claims comparing to previously verified ones. While It is theoretically possible to do all the storage and computation on the blockchain, say Ethereum \cite{ethereum}, but practically, in that case, the cost of operation will become prohibitively large. \\
In the early stages of development, UniqueID will utilize IPFS or other production-ready decentralized storage solutions to store biometric claims. TrueBit protocol offers a general purpose solution to make computation off-chain at the cost of negligible less decentralization.

While TrueBit protocol can practically mitigate the problem of prohibitive computation costs, it is important to note that the special task of checking uniqueness of new biometric claims is capable of being done in parallel. One can think of a smart-contract that offer a reward for the one who finds the closest previously verified unique biometric to the new claim, the results can be verified using a TrueBit like protocol. This design prevents waste of computational resources to a significant level.\\
To get most out of UniqueID network, it is reasonable to develop an independent blockchain and storage layer. It will enable the potential for Proof-of-unique-Human mining system which is explained in ``Novel Mining System" section. 
\subsection{Identity Theft or Death}
Identity needs to be secure against theft or getting lost, on the other hand, to enable some features like UBI, it is required to eliminate dead identities.\\
As we suggested in ``Problem Specification" section, there is a close relationship between the definition of human in our design and trust, based on that point of view we propose a solution for identity theft problem. Each user has to declare a group of at-least 5 trusted ones (friends, family or lawyer) as her trust circle, which have the authority to agree on identity theft or loss and owner of that account can reclaim his identity with new private-key by participating in the validation process.\\
Based on how much the system can tolerate the dead identities, one can think of an extending process for identity holders, similar to new identity registration process with less redundancy.
\subsection{Centralization of Validation Mechanism}
There is always a fear about centralization of processes in decentralized systems, like what happens in Bitcoin mining. The root of this centralization is the economics of scale: the bigger you are, the less you can pay for constant costs and more you can invest in lowering other costs. This can happen for the verification process, and a few verifiers take control over all process. First, because verifiers in disjoint cities cannot be merged and there is no direct competition between them, there is always some geographical decentralization of verifiers. For one city, the hierarchical voting system can prevent such centralization, because it collects meaningful votes, means that you put your vote on someone who you know him, preventing some ``celebrity" or ``briber" from collecting votes. Also, nature of verification process is not something like mining and does not need complex knowledge or expensive devices, and everyone can get into it (It just need your ``time", biometric device and small place). 

\subsection{Fairness}
\subsection{Lack of Incentive}

\section{Analysis}

\section{Applications}

\subsection{Novel mining system}
Satoshi designed a mining system which he called it \textit{one-CPU-one-vote}. The idea behind this is to make it near \textit{one-person-one-vote} because almost everyone has one CPU, and the system remains decentralized because CPU power is distributed around the world. Recently introduced Proof-of-Stake systems have a notion of reducing the energy needed by the proof-of-work system. Both PoS and PoW systems are vulnerable to centralization and other attacks. In UniqueID, however, we have new design space for consensus algorithms.

These algorithms are based on the one-person-one-account property of the system. Having a reliable account on who can enter the system, one can design fair and robust algorithms. Even some naive ideas like Delegated Proof-Of-Stake can change their voting system and become based on real humans, instead of stakeholders. Reputation mining can be designed where everyone has a permanent account. Nothing-at-stake problems can also be mitigated by assigning reputations, such that working on other forks make your reputation near zero. We think that designing efficient, fair and decentralized consensus protocols based on a one-human-one-vote basis is an interesting and independent challenge.

\subsection{Accountable Voting and Survey}
Voting is a critical part of all democratic systems, also surveys play an important role in media and decision making processes, but there's no solution for running a public and accountable online voting without trusting any third-party to protect it against Sybil attacks or fake identities.

Recent advancements in cryptography and blockchains technology offer many powerful solutions for end-to-end secure voting and auditable decentralized voting systems. UniqueID as a decentralized identity solution will make it possible to use all those advancements for social benefits without compromising people privacy.

\subsection{Universal Basic Income}
Universal basic income (UBI) is a kind of welfare system that is based on giving everyone a guaranteed basic income, totally independent of any other income. The idea of national basic income dates back to 18th century but various technical and political issues such as lack of trusted computational infrastructure or a form of trusted universal identity prevented implementation of a UBI system. Though UBI idea originally was presented as a way to save the society from poverty and injustice, many AI experts claim that given the fast and tremendous success of AI technology the world is heading toward a mass unemployment and a UBI system is inevitable and certainly required.

UniqueID provides a trustable universal identity and building a UBI system on top of it is just as easy as writing a simple smart-contract on Ethereum network or any other blockchain with access to UniqueID's validated identities. 

\subsection{Reputation Systems}
Reputation systems are tools to build trust in online communities. Many motivations including the emergence of sharing economy and online marketplaces encourage using a reputation system. Theoretical studies in game theory and practical experiments agree that a valid and reliable reputation system must have a long life length and be protected against Sybil attacks. \\
Creating reputation systems on top of a decentralized identity system such as UniqueID will make Sybil attacks impossible or very expensive and on the other hand, can guarantee a long lifetime.

\subsection{General Framework for Organization}
In an abstract and conceptual level, Bitcoin's idea is to decentralize a specific computational process using redundancy of computation, Ethereum extended this idea and made it possible to decentralize any computational process.\\ UniqueID's idea is to decentralize operation of a specific organization which is responsible for providing universal unique identity, and we claim that by extending same governance scheme and redundancy of operation, one can think of a general framework to decentralize operation of organizations. 

\subsection{Rethinking Governance}
Recently there were a number of discussions about various models of blockchain governance. These discussions mainly focus on pros and cons of on-chain votings with the proof-of-stake mechanism. It seems that these votings can have harmful outcomes, mainly due to the large asymmetry in the number of coins each member of the crypto community holds. A one-person-one-vote system can enable us new ways of thinking about governance problems in blockchain space. As an example, Futarchy, as proposed and combined with blockchain in \cite{Futarchy}, needs such a one-person-one-account system. 

\subsection{Social Media}
\subsection{Fake News}

\section{Reviews}
We categorize previously existing ideas in this area in two different categories, first decentralized identity management solutions which aim to provide a decentralized infrastructure to host centralized issued identities, including DID, Sovrin, and uPort. The second one is decentralized identity issuance solutions which aim to create or define a new kind of identity that is decentralized from its origin, including Democracy Earth and Bitnation, Proof-of-Personhood.

Our work can be categorized in the later one, decentralized identity issuance systems. We want to emphasize that despite previously known solutions, UniqueID can bring decentralized, permanent, unique identity into reality by utilizing biometric authentication and social smart contracts.

In the following section, we briefly review some of the mentioned schemes and compare them to UniqueID in measures of political and operational decentralization, Sybil attack resistance permanence of issued identities, and proper governance and incentive mechanisms. 

\subsection{uPort}

uPort is a decentralized infrastructure for claiming identities and receiving verification from other parties in the network. One can use uPort to host and digitize her identity documents such as national ID card or driving license and get verified by officials or other people for being the owner of the claimed identity.

 First of all, uPort is not designed for a decentralized originated identity which is well defined. Due to lack of a general identifier in the uPort network, uPort do not provide a direct solution to anti-Sybil attack and enabling a universal basic income system. Though uPort provides the possibility to define decentralized originated identities, there is no intrinsic incentive or mechanism to do it.

\subsection{Sovrin}

Sovrin is a protocol and decentralized app based on its own blockchain, aiming to create a sovereign identity and decentralized trust, it is focused on delivering a kind of identity which is secure, private, and partially provable. In the current design of Sovrin system, a trusted set of operations maintain issuing new identities in a semi-decentralized manner. By utilizing its own token, issuers are economically incentivized to participate in the system, though there’s no governance mechanism to prevent issuing fake identities and Sybil attacks. While Sovrin claims that it will become fully decentralized in future but currently there’s no clear scheme for decentralizing issuing new identities in Sovrin network. 

\subsection{Democracy Earth}
Democracy Earth is going to be a decentralized app mainly concerned with providing a infrastructure for liquid democracy and voting. Naturally, it does need a unique identity scheme to prevent Sybil attacks. Although currently it is proposed that each new user upload a video from herself in a concrete format to prove her uniqueness but there’s no clear scheme with an acceptable error that compares videos and verifier uniqueness of new users. It is also suggested that Democracy Earth utilize other platforms to identify its users.
\subsection{everid}
\subsection{ DID}
\subsection{ID2020}

\appendices
\section{Preliminaries}
\subsection{Bitcoin}
\subsection{Smart Contracts}
\subsection{Decentralized Systems}
\subsection{Cryptography}
\end{document}